\documentclass[prl,twocolumn,amssymb]{revtex4}
\usepackage{graphics}
\usepackage{dcolumn}
\begin{document}
\title{Renormalization of the electron-phonon coupling
       in the one-band Hubbard model}
\author{Erik Koch}
\author{Roland Zeyher}
\affiliation{Max-Planck Institut f\"ur Festk\"orperforschung,
             Heisenbergstra\ss e 1, 70569 Stuttgart, Germany}
\date{\today}
\begin{abstract}
We investigate the effect of electronic correlations on the coupling 
of electrons to Holstein phonons in the one-band Hubbard model.
We calculate the static electron-phonon vertex within linear response
of Kotliar-Ruckenstein slave-bosons in the paramagnetic saddle-point 
approximation. Within this approach 
the on-site Coulomb interaction $U$ strongly suppresses the coupling 
to Holstein phonons at low temperatures.
Moreover the vertex function does {\em not} show 
particularly strong forward scattering. Going to larger
temperatures $kT\sim t$ we find that after an initial decrease 
with $U$, the electron-phonon coupling starts to {\em increase} with $U$,
confirming a recent result of Cerruti, Cappelluti, and Pietronero.
We show that this behavior is related to an unusual reentrant behavior
from a phase separated to a paramagnetic state upon {\em decreasing} the 
temperature.
\end{abstract}
\maketitle

The relevance of phonons for high-temperature superconductivity 
has been debated since the discovery of the high-T$_c$ cuprates.
Recently, strong renormalization effects of the electrons near the Fermi
surface, observed in angle-resolved photoemission in several cuprates,
have been at least partially ascribed to phonons \cite{lanzara}. 
Furthermore, quantum Monte
Carlo simulations of the Hubbard-Holstein model suggest that the
electron-phonon coupling shows forward scattering and no substantial
suppression at large U and small dopings \cite{hanke}, 
similar as in the $1/N$ expansion for
the t-J model \cite{zeyher}. On the other hand, it has been pointed 
out \cite{keller} that at small dopings the
Kotliar-Ruckenstein (K-R) slave-boson approach \cite{KR} might yield 
results quite different 
from the 1/N expansion. Below we study the
influence of strong electronic correlations on the electron-phonon
coupling using the K-R approach. The quantity of interest is the static
vertex function $\Gamma$ which acts as a momentum-dependent,
multiplicative renormalization factor for the bare electron-phonon 
coupling.

We consider the one-band Hubbard model on a square lattice with
nearest and next-nearest neighbor hopping, $t$ and $t'$, respectively,
\begin{equation}\label{H0}
 H=-t   \!\!\sum_{\langle i,j\rangle,\sigma}\!
        c_{j\sigma}^\dagger c_{i\sigma}^{\phantom{\dagger}}
   -t'\!\!\!\sum_{\langle\langle i,j\rangle\rangle,\sigma}\!\!
        c_{j\sigma}^\dagger c_{i\sigma}^{\phantom{\dagger}}
   + U \sum_i n_{i\uparrow} n_{i\downarrow}\,.
\end{equation}
For the non-interacting system the dispersion relation is
$\varepsilon_k=-2t(\cos(k_x)\!+\!\cos(k_y))-4t'\cos(k_x)\cos(k_y)$,
and the density of states has a logarithmic van Hove singularity at $4t'$.
For this model we want to study the influence of electronic correlations
on the coupling of electrons to an external field $V_i$. The bare
coupling has the form
$H' = \,\sum_{i,\sigma} n_{i\sigma}V_i$.
Writing $V_i=g u_i$, $H'$ describes also the interaction of electrons and
atomic displacements $u_i$ with coupling constant $g$.
The linear change in the one-particle Green's function
$G(p)$ due to $V_q$ is, 
\begin{equation}\label{var}
{{\delta G(p)}\over {\delta V_q}} = G(p) \Gamma(p,q) G(p+q),
\end{equation}
with the charge or electron-phonon vertex
$\Gamma(p,q) = -\delta G^{-1}(p) / \delta V_q$.
The components of the three-dimen\-sional vectors $p$ and $q$ consist of
a frequency and a two-dimensional momentum.   
For the calculation of the 
vertex 
we use the slave-boson technique of Kotliar and Ruckenstein \cite{KR}.
The basic 
idea of our approach \cite{sblin} is to calculate linear responses by linearizing the 
saddle-point equations for the perturbed system about the homogeneous 
saddle-point solution. We consider only paramagnetic 
solutions. Then there are three slave-bosons, $e$, $p$, and $d$, describing 
empty, singly, and doubly occupied sites, and two Lagrange parameters 
$\lambda^{(1)}$ and $\lambda^{(2)}$ enforcing consistency between 
slave-fermions and slave-bosons. The linear response to a charge-like 
perturbation of a given wave-vector can be determined by solving the 
$5\times5$ system of linear equations given in \cite{sblin}. 
Considering only static fields
$V_{\bf q}$ we find 
\begin{equation}\label{SBvertex}
\Gamma({\bf p},{\bf q}) = 1+{\delta\lambda^{(2)}\over \delta\,V_{\bf q}}
  +z\,\left(\varepsilon_{\bf p}
              +\varepsilon_{\bf p+q}\right){{\delta\,z}\over 
\delta\,V_{\bf q}}.
\end{equation}  
The first term in Eq.~(\ref{SBvertex}) is due to the explicit
dependence of $G^{-1}$ on $V$, the remaining terms are obtained by taking
the derivative of the self-energy with respect to $V$. $\Gamma$ does not 
depend on frequencies because we assumed zero frequency in $q$ and because the
saddle-point self-energy is frequency-independent. 
$z$ is given by the Kotliar-Ruckenstein choice
\begin{equation}\label{KRz}
 z={(e+d)p\over\sqrt{1-p^2-d^2}\sqrt{1-e^2-p^2}}\;.
\end{equation}

In the limit $U\to\infty$ our approach reduces to method (II) of 
Ref.~\cite{keller}, and we have checked that for large $U$ we recover the
results given in their Fig.~1. 

\begin{figure}
 \centerline{\resizebox{2.4in}{!}{\includegraphics{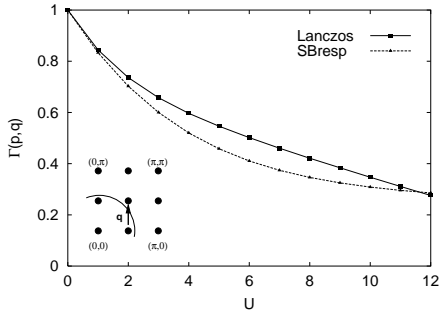}}}
 \caption[]{\label{lanc}
  Electron-phonon vertex $\Gamma({\bf p},{\bf q})$ for the scattering
  of an electron with ${\bf p}=(\pi/2,0)$ by a static phonon of wave-vector
  ${\bf q}=(0,\pi/2)$ (see inset, where the points denote the allowed
  ${\bf p}$ vectors and the solid line the Fermi line) for the Hubbard 
  model with $t'=0$ on a
  $4\times4$ lattice with periodic boundary conditions and 5 up and 5 down
  electrons. The full line gives the result of a Lanczos calculation (with
  fixed number of electrons), the dashed line the results of the slave-boson
  linear response calculation for the same lattice.}
\end{figure}

While it was shown that the slave-boson linear-response method gives very good
results for the charge susceptibility (see, e.g., Fig.~1 of Ref.~\cite{sblin} 
for a comparison with exact diagonalization), it is not clear {\em a priori}
how well it will work for the charge vertex $\Gamma({\bf p},{\bf q})$. 
As a check we have calculated the static vertex 
for a small system using exact diagonalization.
The result for the scattering of an electron from a state 
just below, to a state 
just above the Fermi 
surface is shown in Fig.~\ref{lanc}. Considering the fact that in exact
diagonalization the number of particles is fixed, while the slave-boson
calculations are performed in the grand canonical ensemble, 
the agreement between both
methods is remarkably good. This indicates that the slave-boson 
approach should work well at zero temperature.

\begin{figure}
 \centerline{\resizebox{2.4in}{!}{\includegraphics{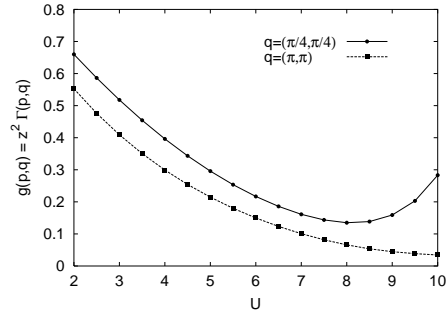}}}
 \caption[]{\label{hanke}
  Effective electron-phonon coupling for the Hubbard model with $t'=0$ on an
  $8\times8$ lattice with filling $n=0.875$ ($28+28$ electrons) as calculated
  in slave-boson linear response at an inverse temperature of $\beta=2$ and for
  electron momenta on the iso-energy-line $\varepsilon_p=0$ (non-interacting
  Fermi surface for half-filling). As follows from Eq.~(\ref{SBvertex}),
  the vertex in slave-boson linear response is then independent of $\bf p$.
  Thus both plots in Fig.~4 of Ref.~\cite{hanke} can be compared to the curves
  shown above. We note that the two calculations differ slightly in the chosen
  filling and in that the QMC calculations of Ref.~\cite{hanke} have not been
  done at $\omega=0$.}
\end{figure}

To find out how well the slave-bosons work at finite temperatures we compare 
to the quantum Monte Carlo (QMC) calculations of Ref.~\cite{hanke}.
Fig.~4 of that work shows the effective electron-phonon coupling 
$g({\bf p},{\bf q})$, as defined in their Eq.~(7), for the Hubbard model on an
$8\times8$ lattice with filling $n=0.88$, calculated at the lowest fermionic
Matsubara frequency, for an inverse temperature of $\beta=2$. For comparison,
we show in Fig.~\ref{hanke} the results of our slave-boson calculations for 
the same model at $\omega=0$ and a slightly different filling $n=0.875$. 
Also here we find a remarkable agreement. In particular, we find that 
after an initial decrease the coupling
for forward scattering (small $\bf q$) starts to {\em increase} for 
$U\gtrsim8$. This seems to indicate that the slave-boson method
also works well at finite temperatures. Moreover, Eq.~(\ref{SBvertex})
naturally explains why the QMC results for 
different electron momenta $\bf p$, shown in Fig.~4 of Ref.~\cite{hanke}, 
are so similar.

\begin{figure*}
 \centerline{\resizebox{6in}{!}{\includegraphics{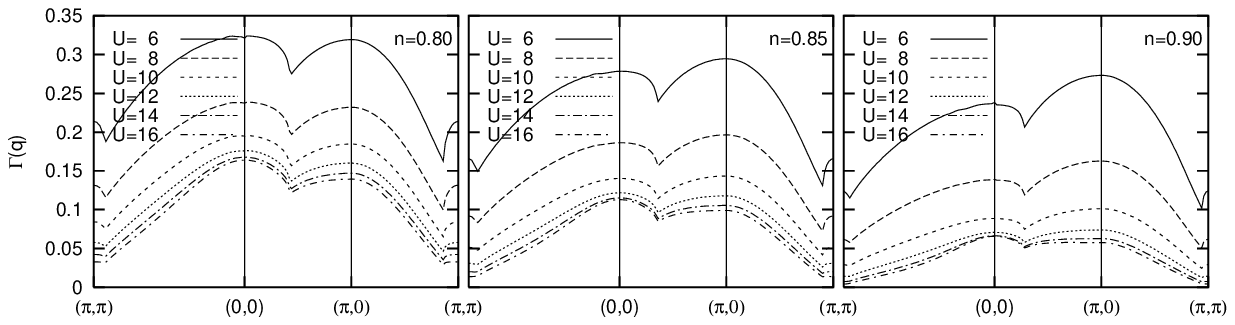}}}
 \vspace{-0.5ex}
 \centerline{\resizebox{6in}{!}{\includegraphics{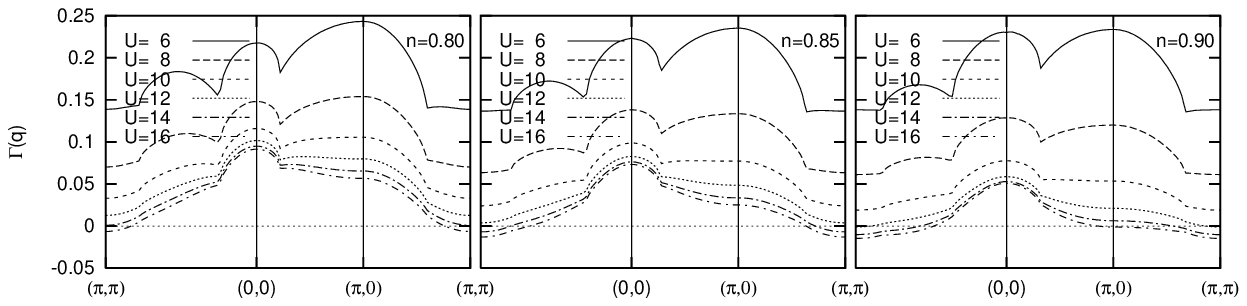}}}
 \caption[]{\label{gamma}
  Electron-phonon vertex $\Gamma({\bf q})$ for scattering electrons on the
  Fermi surface. Calculations are for the Hubbard model on a $1000\times1000$
  lattice with $t'=0$ (top) and $t'=-0.35\,t$ (bottom) at an inverse
  temperature $\beta=500/t$. Fillings and Hubbard $U$ are indicated in the
  plots.  We find that in all cases the vertex function is strongly reduced
  with increasing $U$. For large $U$ the changes become small and eventually
  $\lim_{U\to\infty} \Gamma({\bf q})$ is reached.}
\end{figure*}
After comparing the results of the slave-boson linear response calculations
to more accurate methods, which are, however, limited to small systems
(exact diagonalization) or finite temperatures (quantum Monte Carlo), we
now turn to very large systems at very low temperatures. First we calculate
the electron-phonon vertex for electrons on the Fermi surface, where 
in Eq.~(\ref{SBvertex}) $\varepsilon_{\bf p}$ and 
$\varepsilon_{{\bf p}+{\bf q}}$ are both replaced by the Fermi energy of the
non-interacting system.
Fig.~\ref{gamma}
shows the vertex for momentum transfer $\bf q$ along high symmetry lines
in the Brillouin zone for the Hubbard model 
at essentially zero temperature. The effect of next-nearest neighbor 
hopping is illustrated by comparing calculations for $t'=0$ and $t'=-0.35\,t$.
We find that in both cases the on-site Coulomb interaction strongly reduces
the electron-phonon coupling. This is not completely unexpected as the charge 
response should be strongly suppressed by an on-site Coulomb interaction.
It is, however, in striking difference to the behavior at higher temperature
(Fig.~\ref{hanke}). 
Also, while $\Gamma({\bf q})$ shows a broad peak around ${\bf q}=0$, we 
do not find a particularly pronounced forward scattering. In fact, the 
electron-phonon vertex is often strongest close to ${\bf q}=(\pi,0)$.
This is different from what was found within an $1/N$ expansion \cite{zeyher}.
The $1/N$ expansion relies on the smallness of $1/\delta N$,
i.e., it breaks down at small dopings $\delta$. This can be seen from
the fact that the charge-charge correlation function remains in 
leading order finite for $\delta \rightarrow 0$
though the exact correlation function vanishes in this limit. The
Kotliar-Ruckenstein method, on the other hand, reproduces this limit
correctly in leading order which makes it plausible that in this case
the charge vertex is smaller than in the 1/N expansion, especially
at smaller dopings \cite{keller}. Which of the two methods is more reliable,
in particular, near optimal doping, is not clear and can probably
only be judged by comparison with exact numerical methods.
 
\begin{figure}
 \centerline{\resizebox{2.7in}{!}{\includegraphics{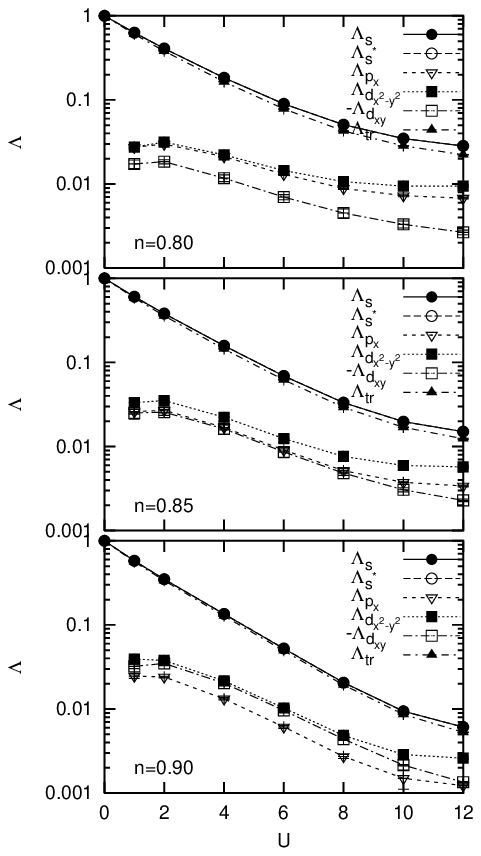}}}
 \caption[]{\label{lambda}
  Renormalization constants $\Lambda_\alpha$ for different pairing channels and
  $\Lambda_\mathrm{tr}$ relevant for transport for the Hubbard model 
  with $t'=0$.
  Calculations were performed for lattices of increasing size at decreasing
  temperatures, performing the $\bf p$ integrals over the whole Brillouin zone
  and weighting with minus the derivative of the Fermi-Dirac distribution 
  $-f'(\varepsilon_p)=
  \beta\,f(\varepsilon_p)\,(1-f(\varepsilon_p))$,
  which becomes a delta functions for $T=0$.
  Convergence of the $T\to0$ extrapolation has been checked by comparing
  $\Lambda_s$ and $\Lambda_{s^\ast}$, which only for $T=0$ are equal. 
  Error-bars for the extrapolation are plotted, but are usually smaller than
  the size of the plotting symbols.}
\end{figure}
To assess the importance of the electron-phonon coupling for superconductivity
we calculate the renormalization factor 
\begin{displaymath}
 \Lambda_\alpha =
  {\int_{\mathrm{FS}}{dp \over|{\bf v_p}   |}
   \int_{\mathrm{FS}}{dp'\over|{\bf v_{p'}}|}
    g_\alpha({\bf p})\Gamma({\bf p},{\bf p'-p})g_\alpha({\bf p'}) \over
   z^2\;
   \int_{\mathrm{FS}}{dp \over|{\bf v_p}   |}
   \int_{\mathrm{FS}}{dp'\over|{\bf v_{p'}}|} g_\alpha^2({\bf p})}
\end{displaymath}
for the pairing channels
$g_s({\bf p})=1$, 
$g_{s^\ast}({\bf p})=\cos(p_x)+\cos(p_y)$,
$g_{p_x}({\bf p})=\sin(p_x)$,
$g_{d_{x^2-y^2}}({\bf p})=\cos(p_x)-\cos(p_y)$, and
$g_{d_{xy}}({\bf p})=\sin(p_x)\sin(p_y)$. 
$\Lambda_\alpha$ is equal to the ratio $\lambda_\alpha/\lambda^{(0)}_\alpha$,
where $\lambda_\alpha$ and $\lambda^{(0)}_\alpha$ denote the
dimensionless electron-phonon coupling constants in the interacting
and non-interacting cases, respectively.
To judge the importance of forward scattering we also calculate
the renormalization factor for transport,
\begin{displaymath}
  \Lambda_{\mathrm{tr}}
 ={\int_{\mathrm{FS}}{dp \over|{\bf v_p}   |}
   \int_{\mathrm{FS}}{dp'\over|{\bf v_{p'}}|}
    \Gamma({\bf p},{\bf p'-p})\,|{\bf v}({\bf p})-{\bf v}({\bf p'})|^2 \over
  2\,z^2\;
   \int_{\mathrm{FS}}{dp \over|{\bf v_p}   |}
   \int_{\mathrm{FS}}{dp'\over|{\bf v_{p'}}|} |{\bf v}({\bf p})|^2}\,.
\end{displaymath}
The results are shown in Fig.~\ref{lambda}. We find that for $U\lesssim10$ 
the $s$-wave couplings decrease almost
exponentially with $U$. For the special case of the Hubbard model with
nearest neighbor hopping only ($t'=0$) we have $\Lambda_s^\ast=\Lambda_s$,
since $g_{s^\ast}$ is constant on the Fermi surface. Moreover 
$\Lambda_\mathrm{tr}\approx\Lambda_s$, reflecting that there is no pronounced
forward scattering; only for larger $U$ does $\Lambda_\mathrm{tr}$ become 
somewhat smaller than $\Lambda_s$. But by then both coupling constants are 
already very small. The higher pairing channels are even weaker, starting from 
zero at $U=0$, going through a maximum around $U=2$ only to decay almost 
exponentially. 
We can thus conclude that within Kotliar-Ruckenstein slave-boson theory,
restricting
the system to be paramagnetic, the contribution of Holstein phonons to 
superconductivity should be very small. 

\begin{figure}
 \centerline{\resizebox{2.5in}{!}{\includegraphics{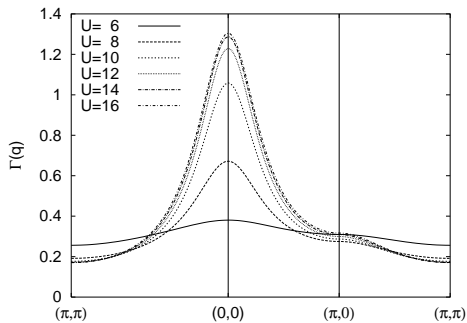}}}
 \caption[]{\label{gammabe}
  Electron-phonon vertex $\Gamma({\bf q})$ for scattering on the Fermi surface.
  Calculations are for the Hubbard model on a $100\times100$ lattice with
  $t'=0$ and filling $n=0.80$ at an inverse temperature $\beta=1$. The plot
  should be compared to the uppermost panel on the right of Fig.~\ref{gamma}.}
\end{figure}

\begin{figure}
 \centerline{\resizebox{2.3in}{!}{\includegraphics{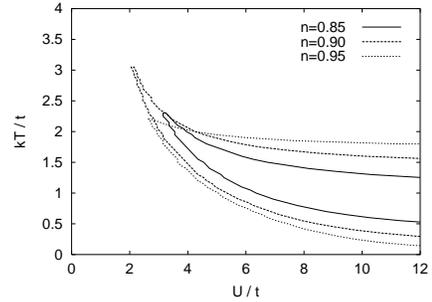}}}
 \caption[]{\label{phasedia}
  Phase separation: The lines enclose the region where the paramagnetic
  slave-boson saddle-point solution is unstable against phase separation.
  Calculations are for the Hubbard model on a $100\times100$ lattice with
  $t'=0$. For $t'/t>0$ the region of phase separation tends to increase,
  for $t'/t<0$ it tends to decrease, in particular, for large doping.}
\end{figure}
We now come back to the surprising upturn of the electron-phonon vertex for
$U\gtrsim8$ shown in Fig.~\ref{hanke}
and also found in the QMC calculations of Ref.~\cite{hanke}. Calculating
$\Gamma({\bf q})$ at $kT\sim t$, we indeed find a drastically different 
behavior than for $T\to0$: Instead of monotonically decreasing with $U$, the 
coupling starts to {\em increase} and develops a very strong forward scattering
peak. An example is shown in Fig.~\ref{gammabe}. Looking at the charge 
response function $\chi({\bf q})$ in the paramagnetic phase we find that this 
behavior is a precursor of a phase-separation instability --- a divergence of 
$\chi({\bf q}=0)$. This has already been pointed out in Ref.~\cite{emmanuele}. 
Calculating the phase diagram, we find a very peculiar reentrant behavior 
around the phase separated region as shown in Fig.~\ref{phasedia}: 
When cooling down the 
system phase separates, but at low enough temperature it reverts back to the 
paramagnetic phase. Since in our calculations we only allow for a paramagnetic
phase, other phases might mask the phase separation. Also, since
slave-bosons may have problems at high temperatures \cite{KR},
it is not clear if the Hubbard model really shows such an reentrant 
behavior. 
Nevertheless, we find a qualitatively similar behavior in the limit 
$U\to\infty$ in the gauge invariant $1/N$ expansion (i.e., a theory without 
Bose condensation).
Phase separation at finite $T$ has also been proposed in Refs.\ 
\cite{Woelfle,Imada,Cappelluti}.
Moreover the good agreement with the quantum 
Monte Carlo calculations of Ref.~\cite{hanke} suggests that our approach might 
indeed capture the relevant physics. It would therefore be interesting to test 
the phase diagram shown in Fig.~\ref{phasedia} with QMC: 
A calculation for, e.g., $\beta=1$ and $U=4\ldots8$, i.e., at temperatures 
and values of $U$, where QMC has little problems, should show clear signs of 
phase separation. Of course, these calculations should be done at $\omega=0$
as the extrapolation from finite Matsubara frequencies might be difficult close
to an instability.

In conclusion, we have studied the influence of strong electronic correlations
on the electron-phonon interaction for the Hubbard-Holstein model
using the Kotliar-Ruckenstein slave boson method. For high temperatures
the boundaries of the phase-separated state were determined in the 
$T-U$ plane for different dopings and the increase of the static vertex 
$\Gamma$ near the boundaries was studied, confirming and extending recent
results of Ref.~\cite{emmanuele}. At low temperatures and moderate or small
dopings we found that $\Gamma$ does not exhibit pronounced forward
scattering behavior and that $\Gamma$ reduces dramatically the 
electron-phonon coupling. It seems that exact numerical 
calculations are necessary to judge the reliability of the 
$1/N$ and the Kotliar-Ruckenstein approaches.
 
We would like to thank O.~Dolgov, O.~Gunnarsson, W.~Hanke, M.~Lavagna,
A.~Muramatsu, and D.~Vollhardt for fruitful discussions.

\end{document}